\documentclass[default]{sn-jnl}

\jyear{2022}%

\theoremstyle{thmstyleone}%
\theoremstyle{thmstyletwo}%

\theoremstyle{thmstylethree}%
\usepackage{amssymb}
\usepackage{multirow}

\begin{document}

\title[MSKACNN for Bearing Fault Diagnosis]{A Multi-size Kernel based Adaptive Convolutional Neural Network for Bearing Fault Diagnosis}


\author[1]{\fnm{Guangwei} \sur{Yu}}

\author[1]{\fnm{Gang} \sur{Li}}

\author[1]{\fnm{Xingtong} \sur{Si}}

\author*[2]{\fnm{Zhuoyuan} \sur{Song}}\email{zsong@hawaii.edu}

\affil[1]{\orgdiv{School of Mechatronic Engineering and Automation}, \orgname{Shanghai University}, \orgaddress{\street{333-9 Nanchen Rd.}, \city{Shanghai}, \postcode{200444}, \country{China}}}

\affil*[2]{\orgdiv{Department of Mechanical Engineering}, \orgname{University of Hawai`i at M\={a}noa}, \orgaddress{\street{2540 Dole St.}, \city{Honolulu}, \postcode{96822}, \state{HI}, \country{USA}}}


\abstract{Bearing fault identification and analysis is an important research area in the field of machinery fault diagnosis. Aiming at the common faults of rolling bearings, we propose a data-driven diagnostic algorithm based on the characteristics of bearing vibrations called multi-size kernel based adaptive convolutional neural network (MSKACNN). Using raw bearing vibration signals as the inputs, MSKACNN provides vibration feature learning and signal classification capabilities to identify and analyze bearing faults. Ball mixing is a ball bearing production quality problem that is difficult to identify using traditional frequency domain analysis methods since it requires high frequency resolutions of the measurement signals and results in a long analyzing time. The proposed MSKACNN is shown to improve the efficiency and accuracy of ball mixing diagnosis. To further demonstrate the effectiveness of MSKACNN in bearing fault identification, a bearing vibration data acquisition system was developed, and vibration signal acquisition was performed on rolling bearings under five different fault conditions including ball mixing. The resulting datasets were used to analyze the performance of our proposed model. To validate the adaptive ability of MSKACNN, fault test data from the Case Western Reserve University Bearing Data Center were also used. Test results show that MSKACNN can identify the different bearing conditions with high accuracy with high generalization ability. We presented an implementation of the MSKACNN as a lightweight module for a real-time bearing fault diagnosis system that is suitable for production.
}




\maketitle

\section{Introduction}\label{sec:intro}
Industrial big data is an important strategic resource for the transformation of the manufacturing industry and an essential element of the Industrial Internet of Things~\cite{Wang:2017}.
At the core of big data is data analysis. To this end, the method of deep learning has become a powerful tool for big data analysis in various fields~\cite{LeCun:2015}.
In recent years, a growing number of research works have been reported in the field of machine fault diagnosis. This work focuses on the utilization of deep learning in identifying and analyzing machine faults that are caused by bearing defects.

Bearings are a key component of rotary machines. 
It has been found that bearing failures are responsible for approximately 30\% of all the failures of rotary machines~\cite{Han:1997}. 
Therefore, bearing condition monitoring and analysis plays an essential role in machine condition diagnosis. 
Traditional methods for diagnosing rotary machine faults include Fourier transform~\cite{Cheng:2018}, empirical mode decomposition~\cite{BenAli:2015}, wavelet transform~\cite{Yu:2017}, singular value decomposition~\cite{Leng:2014}, multi-scale fuzzy entropy analysis~\cite{Zheng:2014}, etc. Recently, with the continuous development of artificial intelligence, convolutional neural networks have achieved great success in the field of pattern recognition~\cite{LeCun:2015}. Fault diagnosis methods have been proposed to perform fault identification based on one or more extracted features found by classic machine learning methods such as support vector machines~\cite{Jiang:2013}, $k$-nearest neighbors classification~\cite{Pandya:2013}, etc. 

Convolutional neural network (CNN) methods proposed by Hinton et al.~\cite{Hinton:2006} and others~\cite{LeCun:2015,Szegedy:2015} have achieved significant success in the field of image and speech recognition. Based on the original CNN, there are some structural variants with superior performance, such as ResNet~\cite{Simonyan:2014}, VGGNet~\cite{HeK:2015}, and LeNet-5~\cite{Wan:2020}. Recently, CNN has been applied to bearing fault diagnosis. Zhang et al.~\cite{Zhangjiangquan:2020} proposed a method based on modified CNNs. The main innovations in their work were the conversion of raw signals into two-dimension images and the verified effectiveness in bearing fault diagnosis. Fu et al.~\cite{Fu:2020} developed a CNN integrated with the adaptive batch normalization algorithm, which uses a large-scale convolution kernel at the grassroots level and a small multi-dimensional convolutional layer. Song and coauthors~\cite{Song:2021} adopted wide kernels of the first two convolutional layers to obtain a larger receptive field based on CNN, and experiments showed that their method performs well in terms of accuracy, noise rejection, and speed. He et al.~\cite{He:2021} proposed a deep transfer learning method based on the one-dimensional CNN (1D-CNN) for bearing fault diagnosis, and adopted correlation alignment to minimize marginal distribution discrepancy between the source domain and target domain, which performs well in comparison experiments.

In bearing diagnosis algorithms, feature extraction based on signal processing is a necessary step~\cite{Brito:2021}.
Traditional methods are based on manual feature extraction~\cite{Wan:2020, Manjurul:2017}, which requires high-level expert knowledge and does not scale to meet the needs of the industrial big data era. 
Most automatic feature extraction methods do not perform well in extracting features that are critical to analyze complex bearing faults such as ball mixing. 
Ball mixing is a common quality issue that occurs in production and assembly proceses of ball bearings. 
Common diagnostic methods are based on frequency-domain analysis, which requires high frequency resolutions in the vibration signals. Increasing the frequency resolutions will cause the sampling time to increase, which further increases the testing time. In addition, frequency-domain analysis tends to result in large and unpredictable errors. Last but not least, the working conditions of bearings are complex. As a result, adaptability of the diagnostic algorithms is a focus of current research~\cite{Zhang:2018439, Yang:2020}.

In this work, we proposed a new bearing fault diagnostic method called multi-size kernel based adaptive convolutional neural network (MSKACNN). 
Major contributions of this paper are summarized as follows:
\begin{enumerate}
	\item[1)] We propose a novel vibration signal diagnostic method based on the 1D-CNN that uses multi-size convolutional kernels in the first layers to improve the feature extraction capability.
	\item[2)] We collected new datasets for ball mixing faults to demonstrate the performance of the proposed method through experiments. MSKACNN produces higher diagnosis accuracy compared with the conventional wide first-layer kernel CNN (WDCNN) and other state-of-the-art alternatives.
	\item[3)] Extensive experiments, including transfer tasks with various loads and machines, are conducted to validate the adaptability of MSKACNN.
	\item[4)] We present a lightweight real-time bearing fault diagnostic software implementation of MSKACNN on an actual bearing vibration data acquisition system to achieve real-time analysis and diagnosis of bearing vibration signals during data collection.
\end{enumerate}

The rest of this paper is structured as follows. A brief introduction to 1D-CNN is provided in Section~\ref{sec:pre}. In Section~\ref{sec:MSKACNN}, the structure of MSKACNN is introduced. Extensive experiments and analyses results are presented in Section~\ref{sec:experiment}. The implementation of MSKACNN as a real-time diagnosis software is discussed in Section~\ref{sec:system}. Finally, Section~\ref{sec:conclusion} concludes the paper.

\section{Preliminaries}\label{sec:pre}
CNN is a multi-stage neural network, including filtering and classification stages~\cite{Krizhevsky:20121097}. The filtering stage is used to extract the features from the input signal while the classification stage classifies the resulting features. The two-stage network parameters are obtained through joint training~\cite{Goodfellow:2016}. The filtering stage consists of basic network elements such as the convolutional layer, pooling layer, and activation layer. The classification stage is generally composed of a fully connected layer. Since bearing vibration signals are one-dimensional, we will focus primarily on 1D-CNN~\cite{Kapoor:2019}.

\subsection{Convolutional Layer}\label{sec:convolution_layer}
\begin{figure}
	\centering
	\includegraphics[width = 0.8\linewidth]{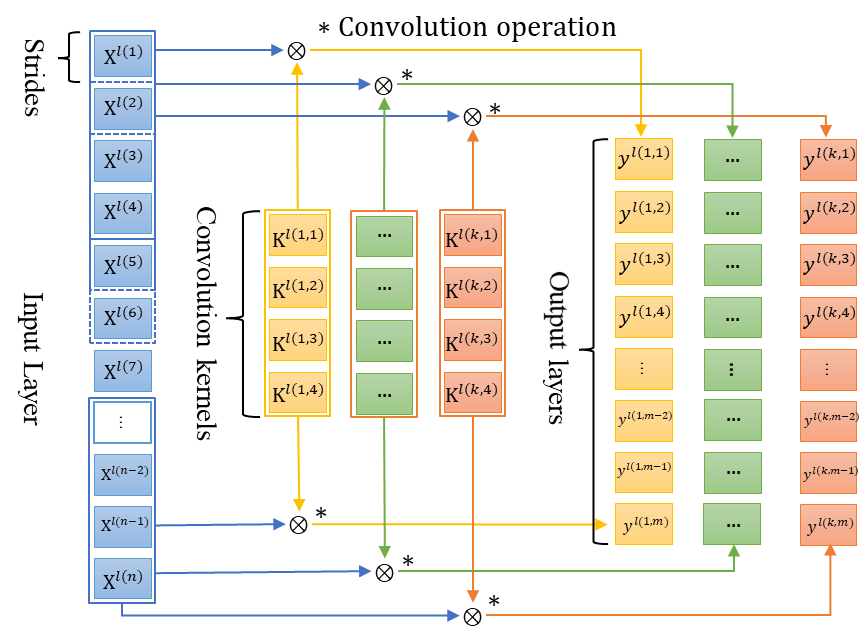}
	\caption{Illustration of the calculation process inside a one-dimensional convolutional layer.}
	\label{fig:1dconvol}
\end{figure}

The convolutional layer uses a convolution kernel to perform convolution operations on local areas of the input signal and generates corresponding features. The most important characteristics of the convolutional layer is weight sharing, that is, the same convolution kernel will traverse the input space once with a fixed step. Weight sharing reduces the network parameters of the convolutional layer and avoids overfitting caused by over-parameterization. The specific convolutional layer operation can be described as 
\begin{equation}
	y^{l(i,j)} = K^l_i * x^{l(r^j)} = \sum^{W-1}_{j'=0}K^{l(j')}_i X^{l(j+j')},
\end{equation}
where $K^{l(j')}_i$ is the $j'$-th weight of the $i$-th convolution kernel for the $l$-th layer, $x^{l(r')}$ is the $j$-th convoluted local area in the $l$-th layer, and $W$ is the width of the convolution kernel. 
The structure of a one-dimensional convolutional layer is shown in Figure~\ref{fig:1dconvol}. There are a total of $k$ convolution kernels, and each convolution kernel traverses the convolutional layer once to perform convolution operations.

\subsection{Activation Function}
\begin{figure}
	\centering
	\includegraphics[width = 0.8\linewidth]{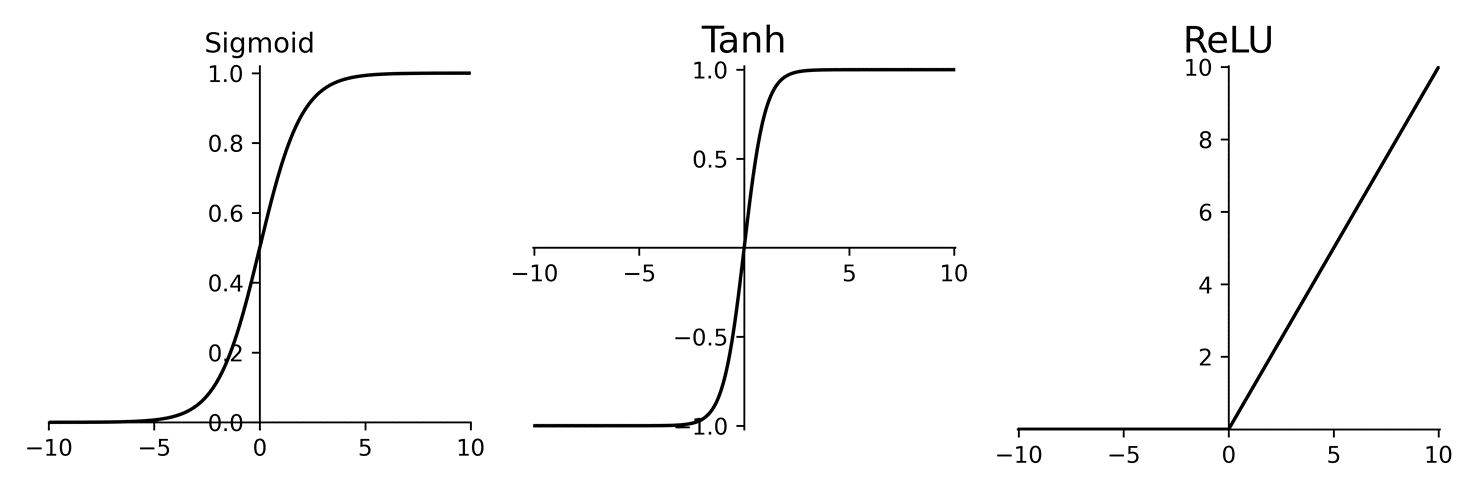}
	\caption{Three commonly used neural network activation functions: Sigmoid, Tanh, and ReLU.}
	\label{fig:activation}
\end{figure}

After convolution, an activation function will perform a nonlinear transformation on each convolution output. The most commonly used activation functions in neural networks design include the sigmoid function, hyperbolic tangent function (Tanh), and rectified linear unit function (ReLU):
\begin{equation}
	a^{l(i,j)} = \textrm{sigmoid}(y^{l(i,j)}) = \frac{1}{1+e^{-y^{l(i,j)}}},
\end{equation}
\begin{equation}
	a^{l(i,j)} = \textrm{Tanh}(y^{l(i,j)}) = \frac{e^{y^{l(i,j)}}-e^{-y^{l(i,j)}}}{e^{y^{l(i,j)}}+e^{-y^{l(i,j)}}},
\end{equation}
\begin{equation}
	a^{l(i,j)} = f(y^{l(i,j)}) = \max(0, y^{l(i,j)}),
\end{equation}
where $a^{l(i,j)}$ is the activation value of convolutional layer output $y^{l(i,j)}$.
Figure~\ref{fig:activation} shows these three common activation functions. Different from the sigmoid and hyperbolic tangent functions, the gradient of the ReLU function is always 1 when the input value is greater than 0, which overcomes the vanishing gradient problem. This paper will use ReLU as the activation function of the CNN.

\subsection{Pooling Layer}
\begin{figure}
	\centering
	\includegraphics[width = 0.8\linewidth]{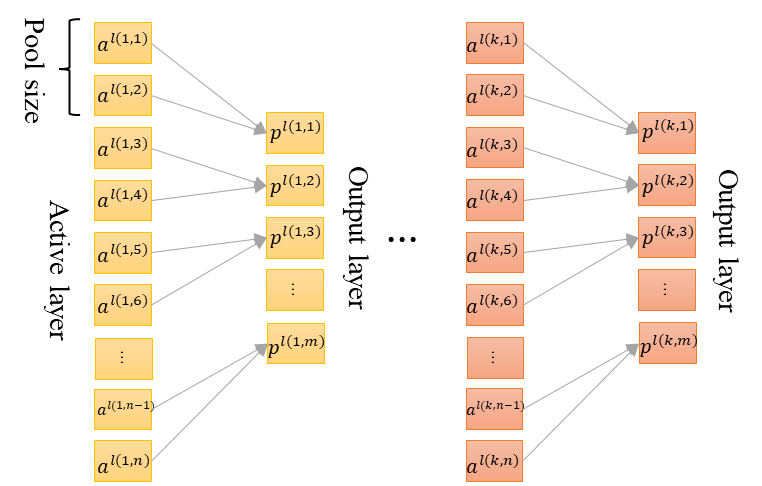}
	\caption{Schematic diagram of one-dimensional pooling operation process.}
	\label{fig:pooling}
\end{figure}

The pooling layer performs a down sampling operation, the main purposes of which are to reduce the number of network parameters, increase the network training speed, and reduce over-fitting. An example of one-dimensional pooling operation is shown in Figure~\ref{fig:pooling}.

Commonly used pooling functions include mean pooling
\begin{equation}
	p^{l(i,j)} = \frac{1}{W}\sum^{jW}_{t \; = \;  (j-1)W+1}a^{l(i,t)},
\end{equation}
and maximum pooling
\begin{equation}
	p^{l(i,j)} = \max\limits_{\{(j-1)W+1 \; < \; t \; \leq \; jW\}} a^{l(i,t)},
\end{equation}
where $a^{l(i,t)}$ is the activation value of the $t$-th neural at the $i$-th frame in the $l$-th layer, $W$ is the width of pooling, and $P^{l(i,j)}$ is the result of pooling layer. In our work, maximum pooling is adopted since maximum pooling can be used to obtain position-independent features for periodic time-domain signals.

\subsection{Fully Connected Layer}
\begin{figure}
	\centering
	\includegraphics[width = 0.8\linewidth]{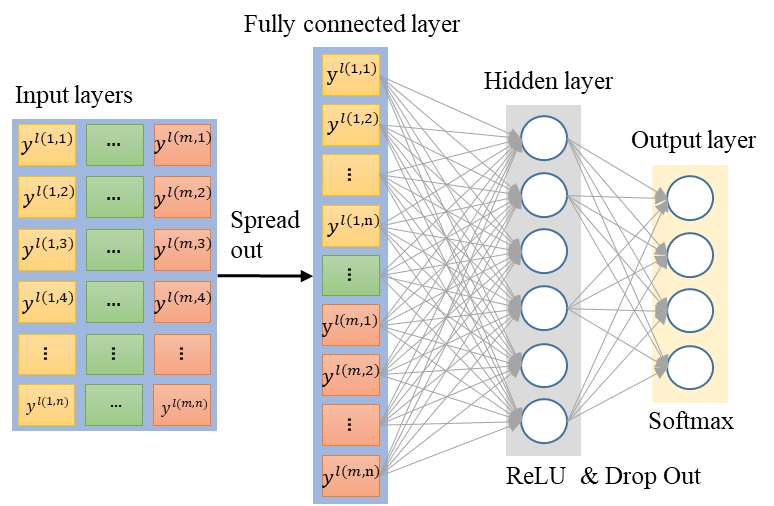}
	\caption{Schematic diagram of the operation of the fully connected one-dimensional convolutional layer and dropout layer.}
	\label{fig:connected}
\end{figure}

The fully connected layer classifies the features extracted from the filtering stage. The specific process is to first stack the outputs of the last pooling layer into a one-dimensional feature vector as the input to a fully connected layer that fully connects the inputs and outputs, as shown in Figure~\ref{fig:connected}. Here, the hidden layer uses the ReLU activation function, and the final output layer uses the softmax activation function.

The forward propagation of the fully connected layer can be described as
\begin{equation}
	Z^{l+1 (j)} = \sum^n_{i=1}W^l_{ij}a^{l(i)} + b^l_j,
\end{equation}
where $W^l_{ij}$ and $b^l_j$ are the weight and offset, respectively, between the $i$-th neuron of the $l$-th layer and the $j$-th neuron of the $(l+1)$-th layer. When the $(l+1)$-th layer is a hidden layer, the activation function is ReLU, and when the $(l+1)$-th layer is output layer, the activation function is softmax, such that
\begin{equation}
	a^{l+1(i)} = \max (0, z^{l+1(j)}),
\end{equation}
\begin{equation}
	a^{j} = \textrm{softmax}(z^{l+1(j)})=\frac{z^{l+1(j)}}{\sum{z^{l+1}}}.
\end{equation}

\subsection{Adaptive Batch Normalization}
Training deep neural networks can be challenging because the distribution of each layer’s inputs changes during the training process, causing an issue known as internal covariate shift. Szegedy and colleagues~\cite{Szegedy:2017} proposed batch normalization to address this by normalizing each dimension of layer inputs over a mini batch re-scale and re-shift the inputs. Batch normalization applies a transformation that maintains a mean output close to 0 and  output standard deviation close to 1.

In intelligent fault diagnosis, distribution divergence often occurs between the training and testing sets, which will lead to a significant degradation in the diagnosis performance of a deep neural network. It is important that batch normalization works differently during training and testing to improve the adaptive ability of model. This is also known as adaptive batch normalization (AdBN)~\cite{Li:2018}.
During training, the layer normalizes its output using the mean and standard deviation of the current batch of inputs. For each channel being normalized, the layer returns 
\begin{equation}
	\alpha \left[\text{batch} - \textrm{mean}(\text{batch}) \right]/ \sqrt{\textrm{var}(batch) + \xi} + \beta,
\end{equation}
where $\xi$ is small constant (configurable as part of the constructor arguments), $alpha$ is a learned scaling factor, and $\beta$ is a learned offset factor. 
During testing, the layer normalizes its output using the averages of the means and standard deviations of the test batches. 

\subsection{Dropout Layer}
The dropout layer randomly sets input units to 0 at each  training step, as shown in Figure~\ref{fig:connected}, which helps prevent overfitting and improves the adaptability of the model.

\section{MSKACNN for Bearing Fault Diagnosis}\label{sec:MSKACNN}
Some bearing states are difficult to identify, such as the mixing ball state, of which features are difficult to extract. Aiming at capturing the detailed temporal characteristics of the bearing, our proposed neural network model adopts large-size convolution kernels of different sizes. This enhances the range of the convolution kernel and allows us to capture both high- and low-frequency features and improve the feature extraction ability. We also add the AdBN and dropout modules to improve the adaptability of the proposed model. In view of the time dependency and periodicity of the bearing vibration signal, the convolution layer is optimized to fully extract the bearing vibration signal features to avoid the loss of temporal characteristics and improve the diagnostic performance of the network. A major structural feature is that the first layer is a multi-size large convolution kernel, and the subsequent convolution layers are all $3\times1$ small convolution kernels. 

\begin{figure}
	\centering
	\includegraphics[width = 1\linewidth]{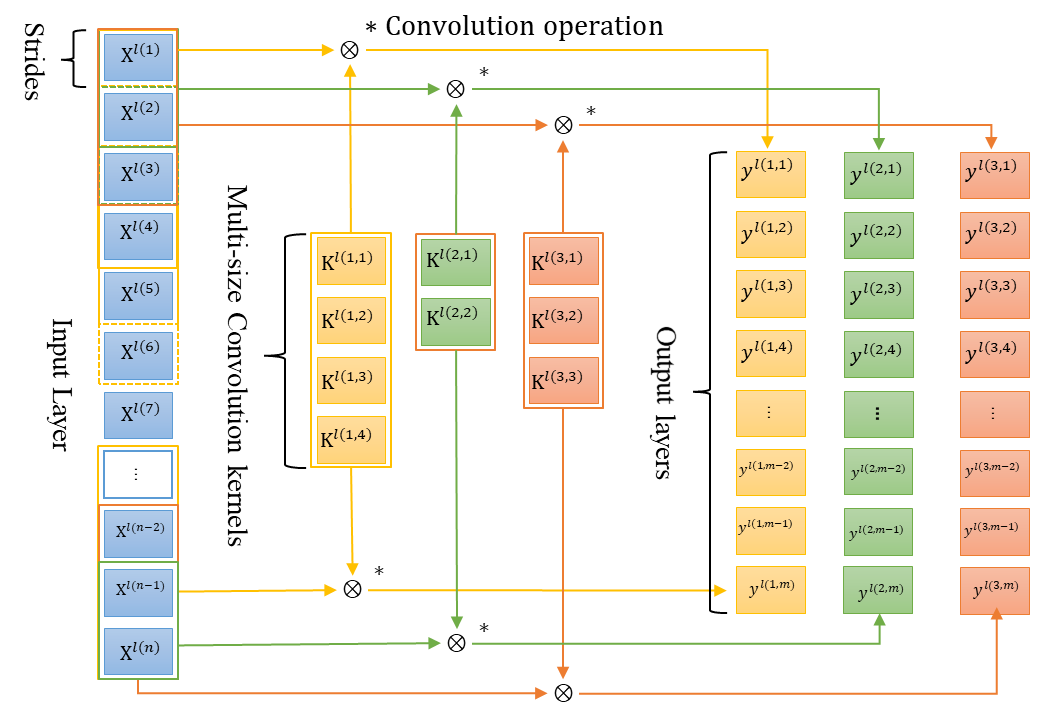}
	\caption{Illustration of the one-dimensional convolutional layer with multi-size convolution kernels.}
	\label{fig:1dconvolmultisize}
\end{figure}

The first layer of MSKACNN is with multi-size convolution kernels (Figure~\ref{fig:1dconvolmultisize}), the purpose of which is to capture the multi-frequency feature information. The goal of applying a large convolution kernel is to extract short-term features, and its function is similar to that of short-term Fourier transform. The difference is that the window function of the short-term Fourier transform is fixed, while the first layer of large convolution kernel is obtained through training. The advantage of CNN is that it can automatically learn diagnostic-oriented features, and automatically remove the diagnostic features that are not informative. Except for the first layer, the size of the convolution kernel of the remaining convolutional layers is $3\times1$. The small number of convolution kernel parameters allows us to deepen the network while suppressing over-fitting. After each layer of convolution operation, adaptive batch normalization is performed. The purpose is to reduce the transfer of internal covariates, increase the network training rate, and enhance the generalization ability of the network.

\subsection{CNN with Multi-size Kernels}
\begin{figure}
	\centering
	\includegraphics[width = 0.8\linewidth]{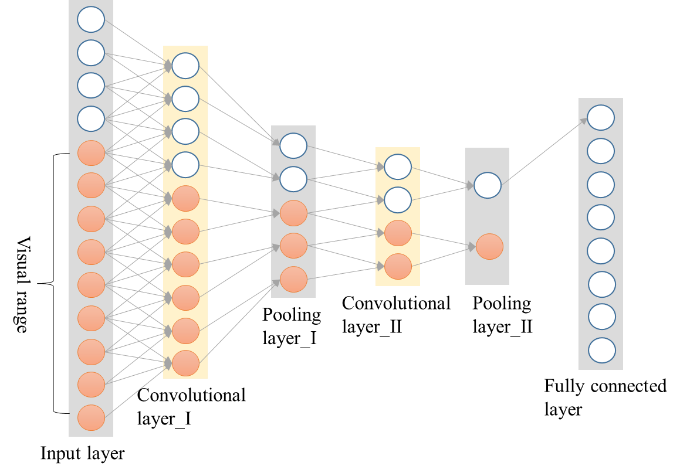}
	\caption{The receptive input field a specific neuron on the last pooling layer.}
	\label{fig:neuronrecoptivefield}
\end{figure}

The selection of the size, step size, and number of convolutional layers is key to the network’s ability to capture key vibration features. The core of the design of a CNN is the receptive field, that is, the perceptive range of a neuron in its underlying network. As shown in Figure~\ref{fig:neuronrecoptivefield}, the inputs range corresponding to the receptive field of a specific neuron in the last pooling layer are marked in red.

Due to the periodicity of the vibration signal, the phase value of each input signal is not necessarily the same. In order for the filter stage of the network to learn displacement-invariant features, the size of the receptive input field of the neurons in the final pooling layer should be greater than one signal period. Assuming that the size of the receptive input field of the neurons in the final pooling layer is $R^{(0)}$, where $N$ is the number of data points collected for one revolution of the bearing and $L$ is the signal length of the input layer, then $N\leq R^{(0)} \leq L$. The relationship between the receptive fields of each layer can be found as
\begin{equation}
	R^{(l-1)} = S^{(l)}(P^{
		(l)}R^{(l)} - 1) + W^{(l)},
\end{equation}
and when $l > 1$, $S^{(l)}=1$, $W^{(l)}=3$, and $P^{(l)}=2$; when $l=n$, $R^{(n)}=1$, leading to
\begin{equation}
	R^{(l-1)} = 2\times R^{(l)} + 2,
\end{equation}
\begin{equation}
	R^{(1)} = 3\times 2^{n-1}  - 2,
\end{equation}
\begin{equation}
	R^{(0)} = S^{(1)}(P^{(1)}R^{(1)}-1) + W^{(1)} = S^{(1)}(3\times 2^n - 5) + W^{(1)},
\end{equation}
where $S^{(l)}$ and $W^{(l)}$ are the step length of the $l$-th convolutional layer and the width of the convolution kernel, respectively, and $P^{(l)}$ is the number of down sampling points in the $l$-th pooling layer. In addition, $L$ needs to be divisible by $S^{(1)}$  such that
\begin{equation}
	L \% S^{(1)} = 0.
\end{equation}

Take the input signal of our experiment later as an example, the input signal length is $L = 4,096$, and the signal period is $N\approx 1,925$. If there are six layers of convolution, then the convolution step length of the first convolution layer can only be 16, and the size of the first convolution kernels $W^{(1)}$ can be 32, 64, 128, and 256.

\subsection{Structure of MSKACNN}
\begin{figure}
	\centering
	\includegraphics[width = 1\linewidth]{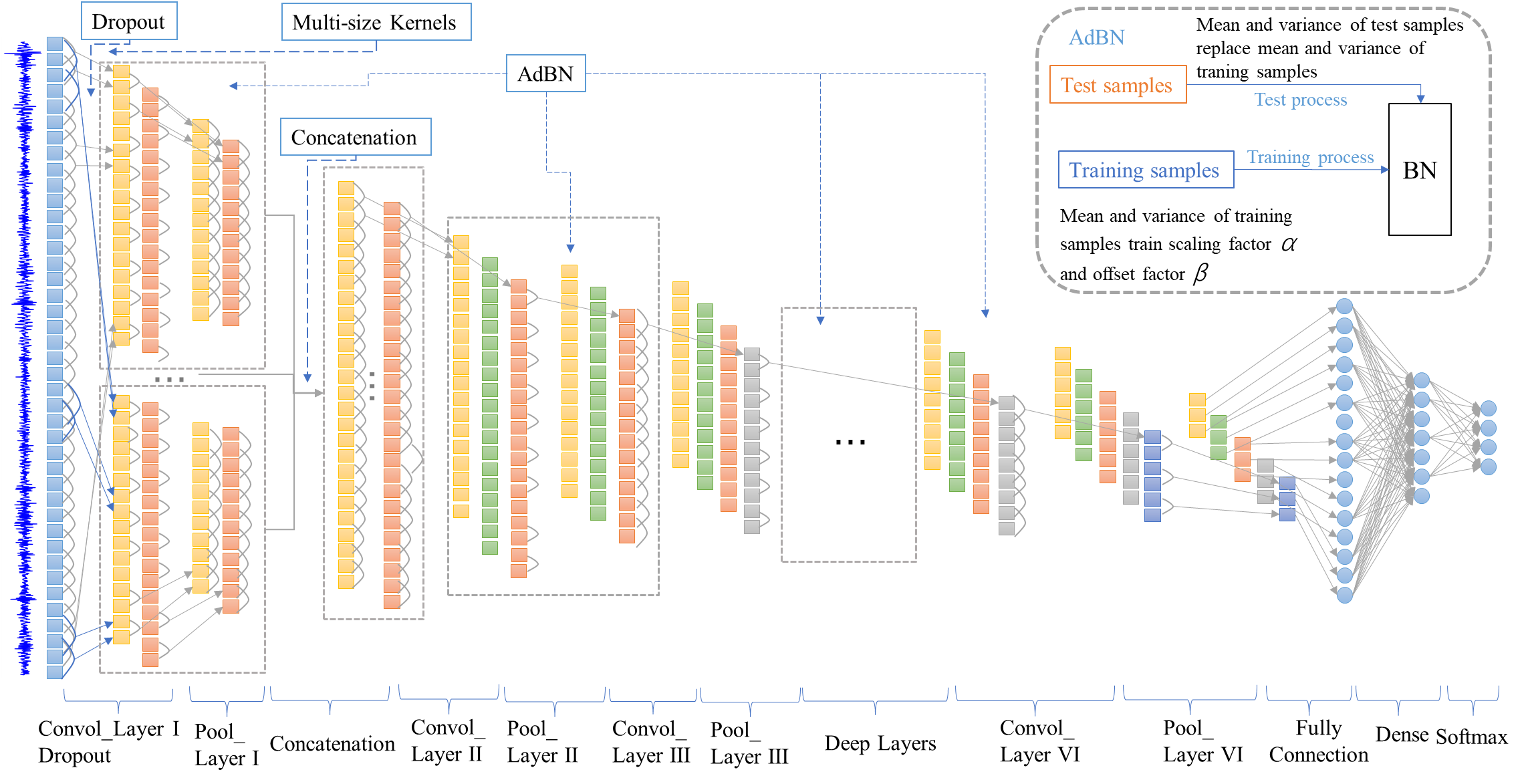}
	\caption{The structural illustration of MSKACNN.}
	\label{fig:mskacnn}
\end{figure}

The complete structure of MSKACNN is shown in Figure~\ref{fig:mskacnn}. The network includes six convolution layers (CV), six AdBN layers (BN), six max-pooling layers (MP), one fully connected layer (FL), one dense layer (DS), one dropout layers, and one softmax layer (SM). The diagnostic signal is activated by the first multi-size convolution layer and becomes a set of feature maps, which are then down-sampled through maximum pooling. This process is repeated before connecting the feature map of a pooling layer to the fully connected layer, where dropout is used in first convolution layer to improve the adaptability of model. The result is then passed to the final softmax layer. The structural parameters of the network are summarized in Table~\ref{tab:param}. The sizes of the convolution kernels of the first layer are $32\times1$, $64\times1$, $128\times1$, and $256\times1$, and the later layers is $3\times1$. The area size of the pooling layer is $2\times1$. Since the first layer is of  multi-size convolution kernels, in order to connect with the subsequent convolution layers, a concatenated layer is introduced here, which takes as input a list of tensors of the same shape except for the concatenation axis and returns a single tensor that is the concatenation of all the inputs.

\begin{table}
\begin{center}
\begin{minipage}{\linewidth}
	\caption{Structural parameters of MSKACNN.}
	\label{tab:param}
	\begin{tabular}{cccc}
		\hline
		Name          & Kernel size / strides                                      & Numbers & Output shape         \\
		\hline
		Dropout       & -                                                          & -       & (4096, 1)            \\
		L1\_CV        & $32\times 1$, $64\times1$, $128\times 1$, $256\times 1/16$ & 4/4/4/4 & (4096, 4) $\times$ 4 \\
		L1\_BN        & -                                                          & -       & (4096, 4) $\times$ 4 \\
		L1\_MP        & $2\times1/2$                                               & -       & (2048, 4) $\times$ 4 \\
		Concatenation & -                                                          & -       & (2048, 16)           \\
		L2\_CV        & $3\times1/1$                                               & 32      & (2048, 32)           \\
		L2\_BN        & -                                                          & -       & (2048, 32)           \\
		L2\_MP        & $2\times 1/2$                                              & -       & (1024, 32)           \\
		L3\_CV        & $3\times 1/1$                                              & 32      & (1024, 32)           \\
		L3\_BN        & -                                                          & -       & (1024, 32)           \\
		L3\_MP        & $2\times 1/2$                                              & -       & (512, 32)            \\
		L4\_CV        & $3\times 1/1$                                              & 64      & (256, 64)            \\
		L4\_BN        & -                                                          & -       & (256, 64)            \\
		L4\_MP        & $2\times 1/2$                                              & -       & (128,64)             \\
		L5\_CV        & $3\times 1/1$                                              & 64      & (128, 64)            \\
		L5\_BN        & -                                                          & -       & (128, 64)            \\
		L5\_MP        & $2\times 1/2$                                              & -       & (64, 64)             \\
		L6\_CV        & $3\times 1/1$                                              & -       & (32, 64)             \\
		L6\_BN        & -                                                          & -       & (32, 64)             \\
		L6\_MP        & $2\times 1/2$                                              & -       & (16, 64)             \\
		L7\_FL        & 64                                                         & -       & (1024, )             \\
		L8\_DS        & 64                                                         & -       & (256, )              \\
		L9\_SM        & 5                                                          & -       & (5, )                 \\
		\hline
	\end{tabular}
\end{minipage}
\end{center}
\end{table}

\section{Experiments and Analysis}\label{sec:experiment}
\subsection{Causes and Effects of Ball Mixing}
Ball mixing, also known as ball size variation, refers to the ball bearing defect where one or multiple steel balls in a bearing are in different dimensions. Because problematic bearing balls are rare and appear in very similar dimensions and colors as normal bearing balls, ball mixing is difficult to identify, causing serious impact on the lifespan, reliability, and safety of the bearing. In the processes of steel ball manufacturing and bearing assembly, the main reasons for ball mixing are loose mounting or damage of steel ball processing, bearing assembly, and auxiliary facilities, resulting in ball jamming, which further leads to ball mixing. Mixed balls can be divided into two types either according to their sizes, i.e., large balls and small balls, or according to their specifications, i.e., homogeneous specifications and heterogeneous specifications, or according to their colors, i.e., mixed dark balls and mixed bright balls.

\begin{figure}
	\centering
	\includegraphics[width = 0.3\linewidth]{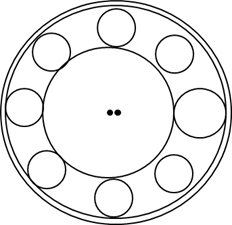}
	\caption{Shaft swing caused by uneven rolling element sizes.}
	\label{fig:shaftswing}
\end{figure}

The uneven size of the rolling elements will cause the bearing axis, therefore the connecting shaft, to swing, as shown in Figure~\ref{fig:shaftswing}. It will also cause variations in the rigidity of the bearing support. The resulting vibration frequencies are $f_c$ and $nf_c \pm f_n$, where $f_C$ is the rotating frequency of the cage and $f_n$ is the rotating frequency of the shaft.

In our test samples, six sets of mixed ball bearings (large mixed balls and small mixed balls) of type 6202 are included. 
The diameter of the bearing steel balls for this model is $6.350~mm$. For the three sets of mixed ball bearings with large balls, one of the seven bearing balls is $1~\mu m$, $2~\mu m$, and $3~\mu m$ larger than the regular balls, respectively. Similarly, for the three sets of mixed ball bearings with small balls, one of the seven bearing balls is $1~\mu m$, $2~\mu m$, and $3~\mu m$ smaller than the regular balls.

\subsection{Data Collection}
\begin{figure}
	\centering
	\includegraphics[width = 1\linewidth]{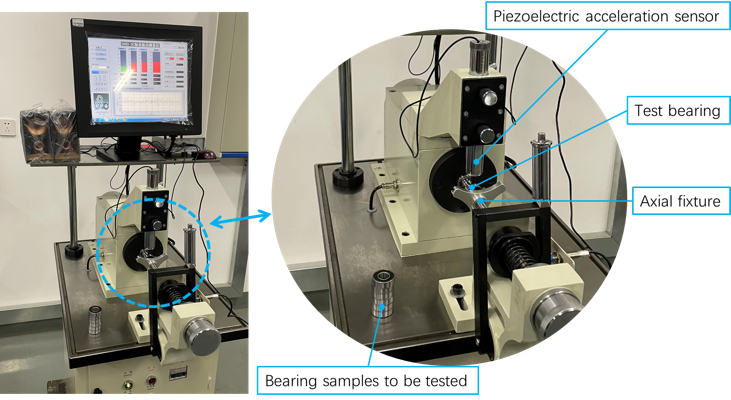}
	\caption{Bearing vibration signal acquisition system.}
	\label{fig:system}
\end{figure}

Data were collected from six types of bearings, namely normal (No), inner ring defect (IR), outer ring defect (OR), rolling element defect or ball fault (Ba), and ball mixing (BM). The defective bearings were manufactured by electrical discharge machining. The data acquisition system is shown in Figure~\ref{fig:system}, including piezoelectric sensors, signal conditioning circuits, motor control systems, and data acquisition cards.

\subsection{Dataset Enhancement}
One of the preferred ways to enhance the generalization ability of machine learning models is to use large training samples~\cite{Goodfellow:2016}. Data augmentation increases the training sample size and diversity to achieve the purpose of enhancing the generalization performance of deep neural networks. In the field of computer vision, dataset enhancement methods include image mirroring, rotation, trimming and scaling, and contrast adjustment~\cite{Krizhevsky:20121097}. In the field of bearing fault diagnosis, however, the one-dimensional vibration signal is periodic with special temporal characteristics, rending the augmentation techniques for image datasets no longer suitable.

\begin{figure}
	\centering
	\includegraphics[width = 0.7\linewidth]{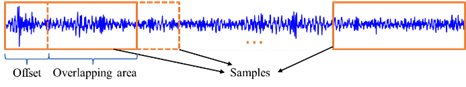}
	\caption{Illustration of the data augmentation method based on overlapping sampling.}
	\label{fig:dataaugmentation}
\end{figure}

For one-dimensional bearing signals, a traditional data augmentation method is overlapping sampling of training samples, where there is an overlap between two consecutive training samples. An example of this sampling method is shown in Figure~\ref{fig:dataaugmentation}. Assuming that a piece of original signal has 491,520 data points, the training sample length is 4,096, and the offset is 487, 1,000 training samples can be obtained, which can well meet the training needs of deep neural networks.

\subsection{Data Collection}
In data collection, the number of data points collected during one shaft revolution can be inferred from the motor speed and the sampling frequency of the sensor. The functional relationship between these parameters can be expressed as
\begin{equation}\label{eq:datapoints}
	N = \frac{60f}{v},
\end{equation}
where $N$ is the number of data points collected per revolution, $f$ is the sampling frequency, and $v$ is the bearing rotating speed. The rotating speed of the bearings during our experiments is 1,500~rpm, and the sampling frequency is 48,128~Hz. Substituting these values into (\ref{eq:datapoints}), the number of data points collected per revolution is calculated to be 1,925. 

In the test, 4,096 data points were diagnosed each time. In order to facilitate the training of the CNN, each segment of the signal $x$ is normalized such that
\begin{equation}
	\tilde{x} = \frac{x - x_{mean}}{x_{std}}.
\end{equation}
A total of five types of bearing fault were prepared for the experiment. The training and test samples were randomly sampled to arrive at a splitting ratio of $4:1$. 

To further evaluate the performance of MSKACNN against other methods, extensive experiments were also conducted on the bearing datasets of from the he Case Western Reserve University (CWRU) Bearing Data Center~\cite{CWRU:2000}. Description of all the datasets as shown in Tables~\ref{tab:dataset} and~\ref{tab:dataset2}. Dataset A (Set A) denotes the data collected with the aforementioned data acquisition system (Figure~\ref{fig:system}) at Shanghai University (SHU).
\begin{table}
	\begin{center}
		\begin{minipage}{\linewidth}
	\caption{Description of the datasets used in the experiments. }
	\label{tab:dataset}
	\begin{tabular}{cccccc}
		\hline
		\multirow{2}{*}{Dataset name} & \multirow{2}{*}{Source} & Fault diameter & Sampling frequency & Speed & Load \\
		&  & \multicolumn{1}{c}{ (mm)} & \multicolumn{1}{c}{(Hz)} & \multicolumn{1}{c}{(rpm)} & \multicolumn{1}{c}{(hp)} \\
		\hline
		Set A & SHU & $\approx 0.5$ & 48,128 & 1,500 & 0.7 \\
		Set B & CWRU & 0.18/0.36/0.53 & 48,000 & 1,772 & 1 \\
		Set C & CWRU & 0.18/0.36/0.53 & 48,000 & 1,730 & 3 \\
		\hline
	\end{tabular}
\end{minipage}
\end{center}
\end{table}

\begin{table}
	\begin{center}
		\begin{minipage}{\linewidth}
	\caption{Sample classes and sample sizes of the data sets.}
	\label{tab:dataset2}
	\begin{tabular}{ccp{0.09\linewidth}p{0.09\linewidth}p{0.09\linewidth}p{0.09\linewidth}p{0.09\linewidth}}
		\hline
		\multirow{2}{*}{} & Classes & Ball fault (Ba) & Inner race fault (IR) & Normal (No) & Outer race fault (OR) & Ball mixing (BM) \\
		Datasets  & Labels & 0 & 1 & 2 & 3 & 4 \\
		\hline
		\multirow{2}{*}{Set A} & Training samples & 2400 & 2400 & 2400 & 2400 & 2400 \\
		& Test samples & 600 & 600 & 600 & 600 & 600 \\
		\hline 
		\multirow{2}{*}{Set B} & Training samples & 2400 & 2400 & 2400 & 2400 & - \\
		& Test samples & 600 & 600 & 600 & 600 & - \\
		\hline 
		\multirow{2}{*}{Set C} & Training samples & 2400 & 2400 & 2400 & 2400 & - \\
		& Test samples & 600 & 600 & 600 & 600 & - \\
		\hline 
	\end{tabular}
\end{minipage}
\end{center}
\end{table}

\subsection{Training and Analysis}
We compare MSKACNN with alternative neural netowrk based methods including the traditional one-dimensional neural network (WDCNN), which is with wide single first-layer kernels~\cite{Zhangwei:2017}, long short-term memory network (LSTM)~\cite{Gao:2021}, and deep neural network (DNN)~\cite{Jia:2016303}. It should be noted that all the methods are performed with AdBN, using the same optimizer, and under the same settings such as the mini-batch size,  condition for early stopping, learning rate, etc. The only differences are in model structures.

The algorithms are implemented using TensorFlow~\cite{Abadi:2016} in Python. The training was performed on a computer with a four-core 4.90 GHz CPU (Intel Core i7-10510U), NVIDIA GPU with 6.1 computing capability, and 16GB of memory. After the training set data pass through the networks, the objective function value is calculated, and then the network weights are updated through the training module. Finally, the trained models are used to diagnose the test samples in the testing stage.

We use the prepared training samples of Set A to train the four aforementioned methods separately. During training, stochastic gradient descent (SGD) with adaptive moment estimation (Adam)~\cite{Kingma:2015} is adopted with a learning rate of $1.0 \times 10^{-4}$. The mini-batch size is 64, and the upper limit of the number of training polls is 2,000. The early-stopping mechanism is adopted, and the training is stopped when the cross-entropy loss value no longer drops significantly within 5 steps.

\begin{figure}
	\centering
	\includegraphics[width = 1\linewidth]{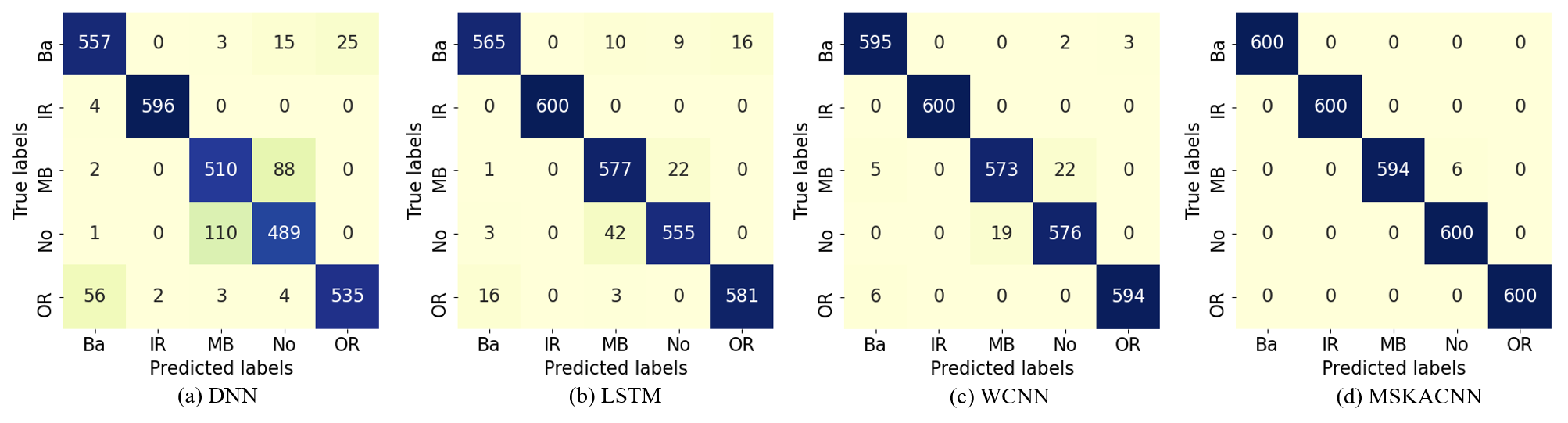}
	\caption{Results on test samples of Set A with DNN, LSTM, WDCNN, and MSKACNN trained on training samples of Set A.}
	\label{fig:confusion}
\end{figure}

\begin{figure}
	\centering
	\includegraphics[width = 1\linewidth]{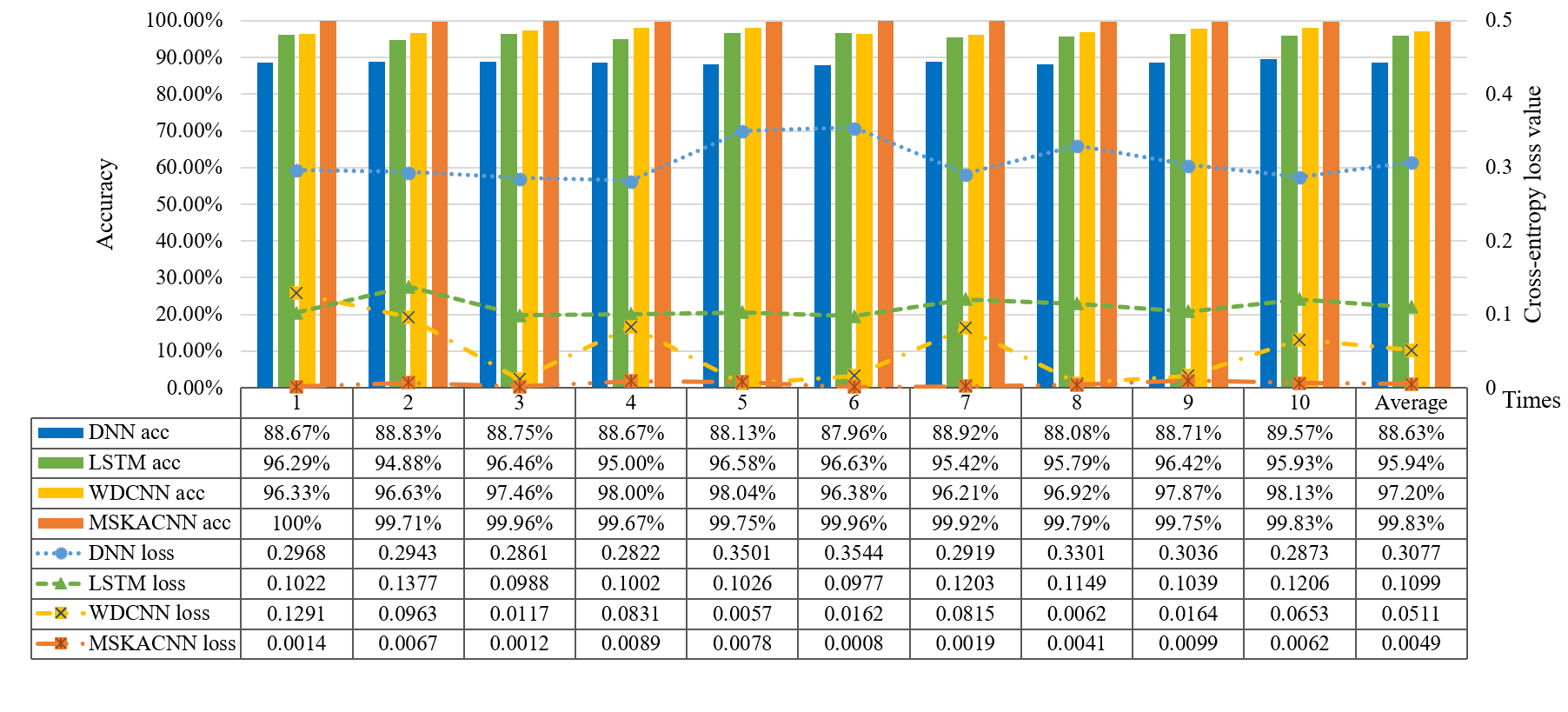}
	\caption{Diagnostic accuracy (acc) and cross-entropy loss value (loss) on the test samples of Set A with DNN, LSTM, WDCNN, and MSKACNN trained on the training samples of Set A.}
	\label{fig:accuracy}
\end{figure}

Since the datasets are limited, in order to avoid randomness, each method was trained 10 times and was tested on Set A each time to record its accuracy. In the Figure~\ref{fig:confusion} shows the performance of four methods on the test samples of Set A. From the heap maps, it can be concluded that the accuracy of MSKACNN is higher than other methods, especially distinguishing ball mixing and normal. The secondly best result was obtained with WDCNN, followed by LSTM and DNN. All four methods achieved good results on identifying IR defects. Figure~\ref{fig:accuracy} shows the accuracy and cross-entropy loss values of each method on Set A. Under the same training conditions, MSKACNN performs better with a lower cross entropy loss value and higher accuracy.

\subsection{Testing with Various Test Datasets}

For intelligent bearing diagnosis, the adaptability and generalizability are important to the diagnostic method. In another word, the model trained on the datasets under one working condition should perform equally well on the test datasets under a different working condition, such as different fault severity, different loads, different vibration measurement systems, etc. In order to verify the generalizability of the proposed model, we prepared datasets of three different working conditions, as shown in Tables~\ref{tab:dataset} and~\ref{tab:dataset2}.

\begin{figure}
	\centering
	\includegraphics[width = 1\linewidth]{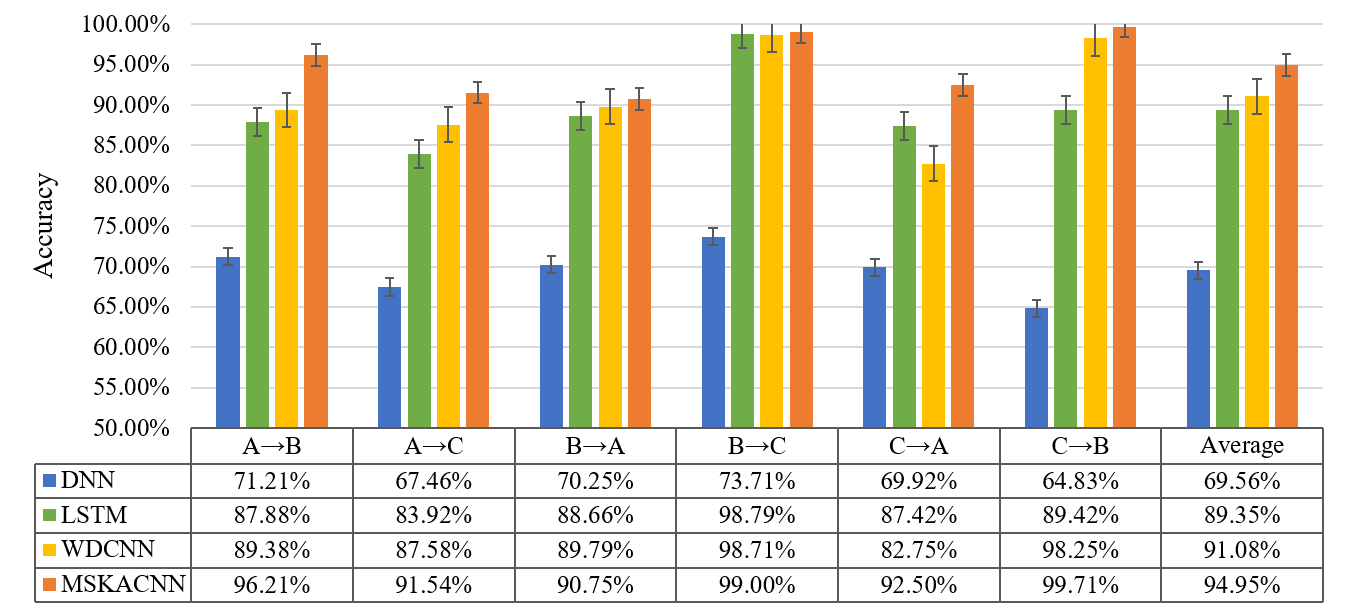}
	\caption{Diagnostic accuracy of transfer diagnosis tasks on different datasets with DNN, LSTM, WDCNN, and MSKACNN. The arrow points from the training sample set to the testing sample set.}
	\label{fig:accuracy4}
\end{figure}

We use dataset names and arrows to indicate the training set and test set used on the the method. For example, task $A\rightarrow B$ denotes that the network is trained by training samples from Set A and tested on test samples from Set B. We call them transfer diagnosis tasks. Each transfer diagnosis task is repeated 10 times to obtain the average accuracy of each method. The diagnosis results of all task are shown in Figure~\ref{fig:accuracy4}.

From the diagnosis results, several key observations can be made. Firstly, MSKACNN outperformed the other methods in most case and achieved a relatively higher average accuracy of all the diagnosis tasks. This indicates great adaptability and generalizability of MSKACNN. Secondly, the diagnosis of each task depends on the distribution divergence between training and testing datasets. For instance, Set B and Set C were collected by the measurement instruments under different operating conditions. As a result, the overall distributions of Sets B and C are rather similar, and tasks B$\rightarrow$C and C$\rightarrow$B often result in the higher accuracy.  However, Set A are different because it was collected with a different set of instrument. Finally, the transfer diagnosis performance may vary slightly with the change of task direction. For instance, the diagnosis accuracy of task A$\rightarrow$B differs from task A$\rightarrow$B for the same method.

\subsection{Feature Visualization}
\begin{figure}
	\centering
	\includegraphics[width = 0.9\linewidth]{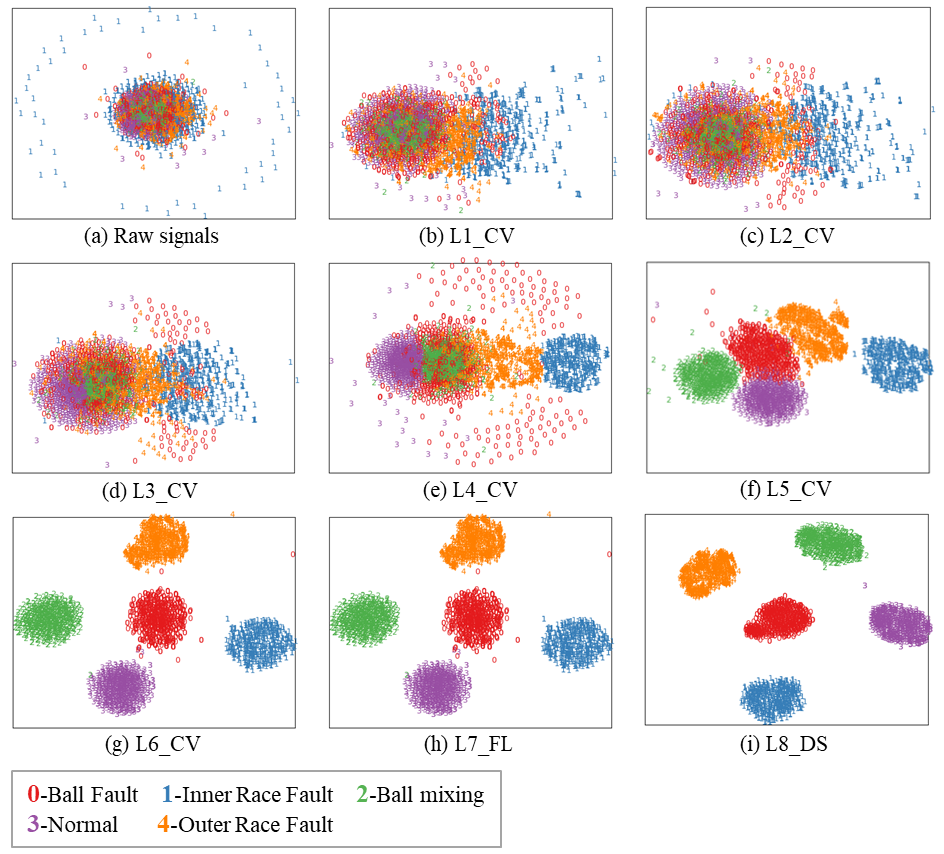}
	\caption{Feature distribution at each layer of MSKACNN on Set A.}
	\label{fig:featuredistriviz}
\end{figure}

To gain more insight about bearing fault diagnosis process and provide a visualization of distribution differences between the diagnosis results of different fault classes intuitively, $t$-distributed stochastic neighbor embedding ($t$-SNE) was utilized to map the features automatically from predicted and target domains into a two-dimensional space. We first visualize the feature distribution obtained MSKACNN on Set A. The features analyzed by all the layers of MSKACNN are shown in Figure~\ref{fig:featuredistriviz}. We note that the first three layers initially try to divide the bearing fault types. The class IF was identified by the fourth layer, and all classes were divided clearly up to the sixth layer, which shows the strong nonlinear mapping capacity of our method.

\begin{figure}
	\centering
	\includegraphics[width = 0.7\linewidth]{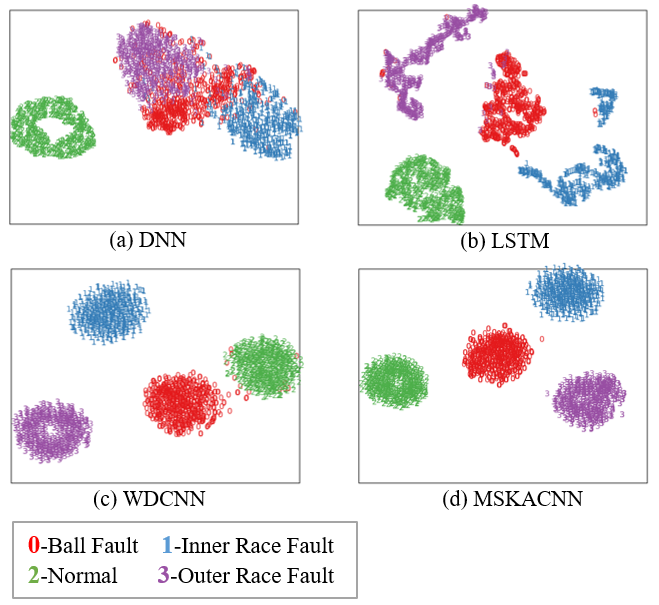}
	\caption{Features distributions in last dense layer of (a) DNN, (b) LSTM, (c) WDCNN, and (d) MSKACNN for transfer diagnosis task A$\rightarrow$B visualized via $t$-SNE.}
	\label{fig:featuredistri}
\end{figure}

Figure~\ref{fig:featuredistri} shows the feature distributions of the four different methods on transfer diagnosis task A$\rightarrow$B.
Samples in the feature space are well separated by MSKACNN (Figure~\ref{fig:featuredistri}d). With WDCNN (Figure~\ref{fig:featuredistri}c), some bearing ball defect (BF) cases were misclassified as normal bearing, which will lead to misidentification of faulty bearings. 
For LSTM (Figure~\ref{fig:featuredistri}b), there are the overlaps between the cases with outer race fault and ball fault. DNN (Figure~\ref{fig:featuredistri}a) performed well for identifying faulty bearings out of normal ones, but does not separate different bearing faults as well as the other methods. It is worth reminding that all methods adopted adaptive batch normalization, which illustrates again that the proposed method has good performance in transfer diagnosis tasks.

\section{Real-Time Diagnostic System}\label{sec:system}
\begin{figure}
	\centering
	\includegraphics[width = 0.5\linewidth]{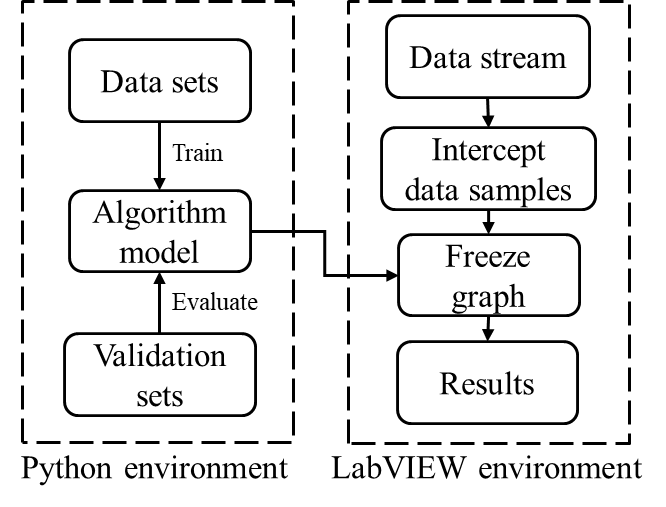}
	\caption{Flow chart of the intelligent real-time bearing fault diagnostic software.}
	\label{fig:diagnoisis}
\end{figure}

It is of economic significance to inspect the bearings either during the production process of the bearings or immediately after their production before them leaving the factory. We embedded the MSKACNN bearing fault diagnostic algorithm into a bearing vibration data acquisition system to achieve real-time online monitoring and identification. The resulting software is lightweight. The implementation process is shown in Figure~\ref{fig:diagnoisis}.

The intelligent diagnostic software was implemented in Python, and the general bearing vibration signal acquisition system was developed in the LabVIEW environment. We converted the built intelligent algorithm model into a frozen graph format, and then used the deep learning module of LabVIEW to embed MSKACNN into the signal acquisition system. During bearing testing process, the vibration signal is intercepted in real-time and MSKACNN is used for real-time diagnosis.

\section{Conclusion}\label{sec:conclusion}
This work focuses on the problem of bearing fault detection and improving diagnostic capabilities for identifying challenging bearing conditions such as ball mixing. We proposed a multi-size kernel based adaptive convolutional neural network (MSKACNN) that directly processes raw bearing vibration signals. The performance of MSKACNN was compared with several alternative methods including WDCNN, LSTM and DNN. A data acquisition system was developed to collect bearing vibration signals from both normal and faulty samples for both training and testing. We also considered the bearing dataset from CWRU Bearing Center in our experimental analysis and demonstrated that MSKACNN has high adaptability in transfer diagnosis tasks. Finally, we presented an implementation of MSKACNN as a real-time bearing diagnosis module in actual bearing production systems.

The accuracy and richness of datasets are key to machine learning algorithms. In the future, we plan to perform experiments with MSKACNN on more complex bearing conditions such as compound faults, and optimize the structure of network model accordingly, including but not limited to developing ensemble learning models for bearing diagnosis.

\bmhead{Acknowledgments}
ZS would like to thank the U.S. National Science Foundation (NSF) for partial funding under award AI Institute in Dynamic Systems (CBET-2112085).

\bibliography{ref}


\end{document}